# Deterministic formation of highly coherent nitrogen-vacancy centers using a focused electron irradiation technique

*Claire A. McLellan[†], Bryan A. Myers[†], Stephan Kraemer[‡], Kenichi Ohno[§,⊥], David D. Awschalom[§], Ania C. Bleszynski Jayich[†]\**

† Department of Physics, University of California, Santa Barbara, Santa Barbara CA 93106 USA

‡ Materials Department, University of California, Santa Barbara, Santa Barbara CA 93106 USA

§ Institute for Molecular Engineering, University of Chicago, Chicago, Illinois 60637 USA

⊥ Current address: Applied Materials, Inc., 3340 Scott Boulevard, Santa Clara, CA, 95054 USA

\* Corresponding Author

Abstract: We demonstrate fully three-dimensional and patterned localization of nitrogen-vacancy (NV) centers in diamond with coherence times in excess of 1 ms. Nitrogen δ-doping during CVD diamond growth vertically confines nitrogen to 4 nm while electron irradiation with a transmission electron microscope (TEM) laterally confines vacancies to less than 1 μm. We characterize the effects of electron energy and dose on NV formation. Importantly, our technique enables the formation of reliably high-quality NV centers inside diamond nanostructures, with applications in quantum information and sensing.

Because of its solid-state host, natural coupling to photons, and long quantum coherence at room temperature, the NV center spin is promising in quantum photonics[1,2], nanometer-scale field sensing[3,4], and hybrid quantum networks[5]. However, for reproducible and reliable scaling, a critical requirement is the deterministic placement of NVs in the optimal location of a diamond nanostructure while preserving the NV's excellent spin properties. For example, the NV may need to be placed in the largest electric field of an optical cavity[6], the largest strain of a mechanical resonator[7], or near the apex of a scanning probe tip[8]. Lateral precision of ~ 10-100 nm and depth precision of ~ 1-10 nm are critical for these applications. Even more aggressive proposals to entangle NV qubits through proximal magnetic dipolar interactions, or to spin chains (dark spins) require three-dimensional precision down to < 10 nm[9,10].

The NV center is a point defect in diamond created by replacing two adjacent lattice sites with a nitrogen and a vacancy[11]. Conventional methods of localizing NV centers inside nanostructures or in close proximity to each other rely on the random statistics of nitrogen and vacancy placement, with density the only tunable parameter. A major drawback of these methods is that they are not scalable; it is necessary to fabricate many devices to form one usable device with an NV in an optimal location[6]. In some cases, it is possible to register the fabrication process around a NV center, but this registration process is limited in precision by optical diffraction, is time consuming, and not scalable[12].

Recently, research groups have explored a variety of patterning methods to place NV centers deterministically. However, many of these methods result in severely compromised NV spin coherence times and/or require post microfabrication processes that are incompatible with fabricated devices. For instance, Toyli *et al.* demonstrated lateral control of NVs by ion implanting through nanofabricated apertures[13], but the resulting coherence times were relatively short (<~ 20 μs). Focused nitrogen-ion[14–16] and helium-ion[17] implantation can also localize NV centers but the resulting coherence times are either short, < 1 us, or not reported. Long NV coherence times (~ 700 μs) were demonstrated in nitrogen δ-doped diamond via carbon implantation through apertures[18]. However, this process requires patterning a mask on the diamond, precluding its use with *e.g.* suspended diamond structures such as cantilevers or photonic crystals.

In this letter, we demonstrate a simple, maskless method to create highly coherent, patterned NV centers using the focused electrons of a transmission electron microscope (TEM) and nitrogen δ-doped diamond. The technique provides precise, lateral control of NV center position, and in conjunction with nitrogen δ-doping achieves full three-dimensional control over NV position at the nanoscale. Furthermore, TEM irradiation is highly tunable and allows for careful studies of the interaction of electron beams with diamond, and the effects of electron energy and dose. Using the NV centers as reporters, we determine the threshold energy for vacancy formation by electrons and displacement energy of carbon in the diamond lattice.

We prepared the samples for this study using a δ-doping technique[19]. Using a plasma-enhanced chemical vapor deposition process, we first grew a thin film (~50 – 80 nm) of $^{12}C$ isotopically pure (99.99%) diamond on an electronic-grade diamond substrate (Element Six), as described in earlier work[19] (Figure 1a). The films were δ-doped with $^{15}N$ during growth. The $^{15}N$ layer is approximately ~ 4 nm thick and the depths of the NVs for the two samples presented are 20 nm and 50 nm, which were set by the thickness of the grown diamond capping layer[19]. Prior to growth, the substrates were polished to an RMS surface roughness of < 1 nm and subsequently etched (500 nm) with ArCl ICP ion etching[20] to remove polishing-induced strain in the diamond. The 20-nm-deep NV sample was used to study the effectiveness of the TEM irradiation to form shallow NVs. The 50-nm-deep NV sample was used to study NV properties without being dominated by surface effects[21].

Vacancies were introduced locally into the diamond via focused electrons from an FEI Technai G2 Sphera TEM at room temperature (fig. 1a). The TEM provides *in situ* tuning of the energy, current, and areal dose of the electron irradiation. We investigate energies ranging from 120 keV to 200 keV, doses ranging from $10^{14}$ e/cm$^2$ to $10^{18}$ e/cm$^2$, and spot sizes ranging from 0.9 - 20 um. After irradiation, the sample was annealed at 850°C in H$_2$/Ar forming gas for 2 hours. To stabilize the NV$^-$ charge state, the surface was oxidized with a boiling acid mixture of nitric, sulfuric and perchloric acids at ~210°C for one hour.

To image and characterize the NVs formed by TEM irradiation, we use a homebuilt confocal microscope. A microwire draped over the diamond surface transmits the radio-frequency pulses used for electron spin resonance measurements. A permanent magnet provides the Zeeman splitting (~ 40 G) of the NV center spin levels.

Figure 1b shows a surface confocal photoluminescence (PL) image of a TEM-irradiated diamond sample containing a 50-nm-deep $^{15}$N δ-doped layer. An optically detected magnetic resonance (ODMR) sweep (fig. 1c) confirms the presence of NV centers in the δ-doped layer, as evidenced by the characteristic $^{15}$N hyperfine coupling. This $^{15}$N signature allows us to distinguish the NVs in the δ-doped layer from the predominantly (99.6%) $^{14}$NV found in the bulk substrate.

A central result of the current work is the reproducible creation of shallow NVs with long coherence times. Figure 2a shows a characteristic plot of NV coherence, as measured with a Hahn echo pulse sequence, for a 50-nm-deep, δ-doped NV that was irradiated with 200-keV electrons at a dose of 2x10$^{18}$ e/cm$^2$. The $T_2$ coherence time extracted from a fit to the data is 1.0 ± 0.15 ms. Figure 2b shows a $T_2$ histogram of measured NVs for the 50-nm-deep sample; all NV centers exhibit $T_2$'s longer than 100 μs, and three have $T_2$'s in excess of 1 ms. The consistently long coherence times highlight the gentle nature of our NV formation technique. The measured NVs sample the full range of our electron irradiation parameters, and there was no correlation observed between $T_2$ and either electron dosage or energy in the ranges we studied.

NV decoherence can be caused by a variety of sources, with common culprits being a $^{13}$C nuclear spin bath, a paramagnetic spin bath of substitutional nitrogen (P1) centers, and other lattice defects often associated with implantation- or irradiation-induced damage. We disregard decoherence from the $^{13}$C nuclear spin bath because of the low concentration (0.01%) of $^{13}$C[22]. To understand what limits $T_2$ to 1 ms in our samples, we consider the limits imposed by a bath of P1 centers. Assuming the theoretical prediction from Wang and Takahashi of $T_2 \sim 20$ (μs ppm)/$\rho_N$ where $\rho_N$ is nitrogen concentration[23], we estimate $\rho_N \sim 20$ ppb in the case that our $T_2$ is in fact limited by nitrogen spins. We note that this theoretical estimate of $\rho_N$ is consistent with previous depth-dependent coherence studies in N δ-doped diamond,[21] accounting here for N to NV conversion. By counting NV centers in an area irradiated with 200-keV electrons at a dose of $2 \times 10^{18}$ e/cm$^2$ (as in fig. 3a) we estimate the $^{15}$NV density ($\rho_{NV}$) in this area to be ~ 14 $^{15}$NVs/μm$^2$, or ~ 20 ppb $^{15}$NV in a ~ 4-nm thick layer. We arrived at this estimate by counting the number of observed $^{15}$NV orientations, a number between 1 and 4, in each diffraction limited spot of a 27 μm$^2$ area and then correcting for the number of $^{15}$NVs through a maximum likelihood estimation.

Hence, the measured $T_2$'s are consistent with limits imposed by a paramagnetic spin bath of P1 centers, indicating that the TEM irradiation introduces minimal lattice damage. In contrast, NV ensembles produced via N ion implantation typically exhibit average $T_2$ times shorter than the P1-imposed limit. Interestingly, we note that our results indicate a relatively high conversion efficiency from N to NV of ~50%.

Our TEM-based technique provides a high degree of control. The density of NVs is highly sensitive to the electron dose, and the depth at which NVs are formed is determined by the energy of the incident electrons. Figure 3a shows confocal photoluminescence images of three areas of a diamond sample subject to different electron doses. The change in photoluminescence (PL) between the images shows that for higher doses the number of created vacancies increases. Figure 3b plots $\rho_{NV}$ as a function of electron irradiation dose, where $\rho_{NV}$ was measured as described earlier. With increasing electron dose, the combination of increasing $\rho_{NV}$ with no observed decrease in $T_2$ is promising for ensemble magnetometry. Furthermore, there is no evidence of saturation in $\rho_{NV}$ and hence in the N-NV conversion in the dosage range studied. Therefore, it is possible that higher electron doses could further enhance magnetic sensitivity.

Depth control of the NV formation is demonstrated in figure 4: irradiating with higher electron energies increases the depth extent to which NVs are formed in the diamond. To quantitatively map out this dependence, the formation of $^{14}$NV in the bulk diamond is used as a marker. The depth of an NV was determined by the microscope's focus height, with aberrations accounted for in a wave-optics model[24]. Optical diffraction and spherical aberrations also limit the measurement precision as the focal plane extends deeper into the diamond. Although the vacancies diffuse during annealing, previous diffusion studies suggest that under our annealing parameters, vacancies only migrate hundreds of nanometers[25,26], a value smaller than the uncertainty in our optical depth measurement. Figure 4 plots the depth of the deepest observed NV, where NVs were formed in a layer that extends from the surface down to this measured depth. A linear fit to the NV depth vs. irradiation energy data yields the minimum threshold energy $E_T$ required to form an NV center right at the surface, $E_T = 145 \pm 16$ keV. The slope of the linear fit is 1.05 ± 0.04 µm/keV, *i.e.* increasing the electron energy by 1 keV results in the formation of NVs 1 µm deeper in the diamond.

Using CASINO V2, a Monte Carlo simulator of electron trajectories in solids[27], we simulated the maximum travel depth of electrons before their energies fall below $E_T$ = 145 keV. The results of the simulation are shown as a green line in figure 4. Though $E_T$ has been fit from the data, the slope of the green line (1.05 ± 0.01 µm/keV) is not an adjustable parameter. The good agreement between data and simulation indicates that NV creation is a good marker for vacancy formation. Using a collision model from Campbell *et al.* we calculate the displacement energy $E_d$ for a carbon to be removed from the diamond lattice to be 30 ± 4 eV, in good agreement with previous studies[28–31].

An attractive feature of TEM irradiation is its facile ability to laterally localize NV centers in a point-and-shoot fashion. To demonstrate this capability, we focused a 200-keV electron beam down to a 0.9-µm-diameter spot size and irradiated an array of 6 spots spaced 3 µm apart with a dose of 3.75x10$^{17}$ e/cm$^2$, as schematically shown in figure 5. Figure 5 also shows a confocal PL scan of the irradiated area after subsequent annealing. The irradiation pattern is faithfully replicated by NV center PL, where each irradiated spot contains ~1-4 NV centers, as confirmed through ODMR measurements. A highly desirable future prospect is to use the TEM's imaging capabilities to locate a pre-patterned feature, such as a diamond nanopillar or photonic crystal cavity, and locally irradiate to form an NV in the desired location. The likelihood of forming a high-quality NV in a nanofabricated device is high because, as demonstrated in this paper, TEM irradiation creates a high density of NVs with long coherence times.

We note that TEM irradiation also forms NVs outside of the irradiated area, but interestingly only at the surface, as determined to within axial optical diffraction. The image in Figure 1b shows surface NVs formed outside the vicinity of the 15-μm-diameter irradiated area; these NVs extend up to hundreds of microns from the edge of the irradiated area. We hypothesize that surface channeling of TEM electrons is at play.

In conclusion we demonstrate a simple, TEM-based method to form patterned, localized NVs with coherence times in excess of 1 ms. Using the focusing capabilities of the TEM, we localize the lateral position of NVs to less than 1 μm, and nitrogen δ-doping provides depth localization to several nanometers. By varying the energy of the electrons we demonstrate control over NV density and depth and determine the threshold electron energy for NV formation to be 145 keV corresponding to a 30-eV displacement energy of carbon in diamond.

The method we have demonstrated here opens the door to several future improvements in NV-based technologies. It will be interesting to explore the upper limits on magnetic sensitivity using high densities of TEM-induced, δ-doped NV centers. Recently, 200-keV TEM irradiation of Type-1b diamond has been used to create NV center densities as high as 7 ppm[32]. It will also be interesting to measure the optical properties of NVs formed via this technique, as its gentle nature is likely to produce NVs with optical linewidths more stable than those typically observed in implanted NVs. Lastly, we look forward to the deterministic positioning of NV centers inside fabricated nanostructures, such as mechanical resonators or photonic crystals, together with pushing the limits of nanoscale placement down to the 10-nm range.

Acknowledgement: This work is supported by the Air Force Office of Scientific Research MURI programme and the DARPA QuASAR program. The research involved the use of shared experimental facilities of the Materials Research Laboratory at UCSB supported by the MRSEC Program of the National Science Foundation under Award No. DMR 1121053. C. A. M acknowledges support by the National Science Foundation Graduate Research Fellowship under Grant No. 1144085. B.A.M. acknowledges support from a Department of Defense fellowship (NDSEG) and an IBM PhD fellowship.

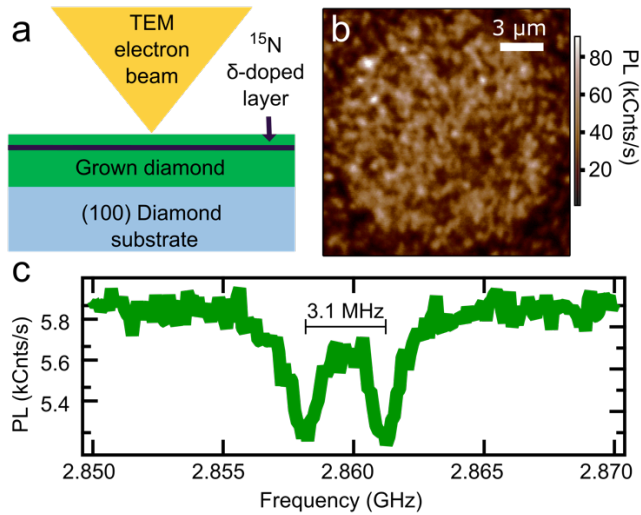

**Figure 1.** (a) Schematic of the experiment: a $^{15}$N δ-doped, CVD-grown diamond is irradiated with a focused TEM electron beam to form localized vacancies. NV centers are formed after subsequent annealing. The 4-nm-thick $^{15}$N layer lies between a 32-nm-thick buffer layer and a 20 (or 50)-nm-thick cap. (b) Confocal photoluminescence image of the surface of a TEM-irradiated area of the diamond showing several individually resolvable NV centers. The area was irradiated with a 15-μm-diameter electron beam, 200-keV incident energy electrons, and a dose of $10^{17}$ e/cm$^2$. This image also shows NVs at the surface outside of the irradiation. (c) Optically detected magnetic resonance signal from a δ-doped $^{15}$NV showing the signature 3-MHz hyperfine peak splitting.

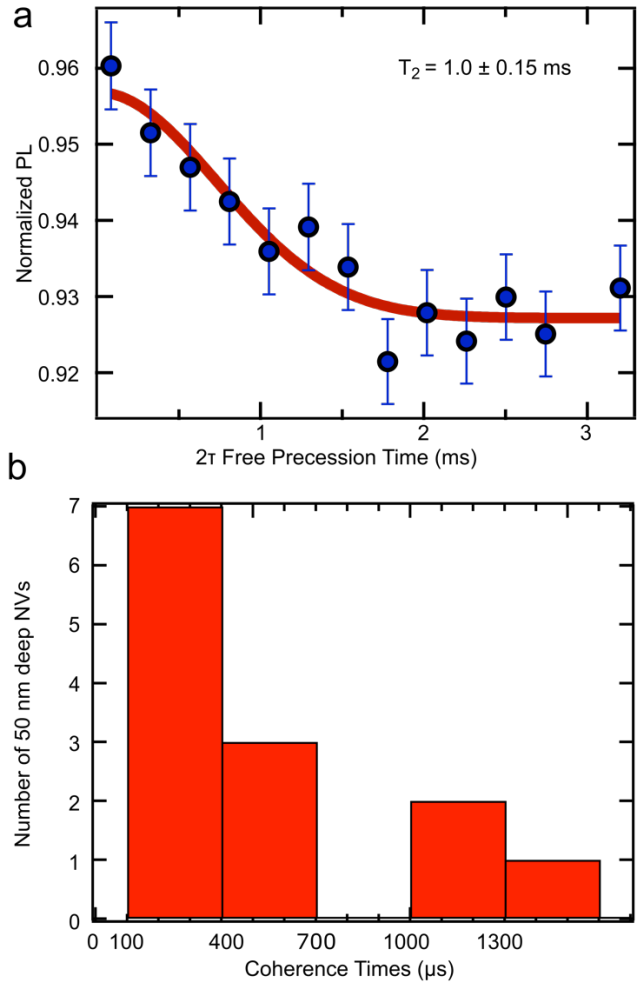

**Figure 2.** (a) A characteristic $T_2$ of a 50-nm-deep NV formed by combining nitrogen δ-doped diamond with TEM irradiation. (b) Histogram of measured $T_2$ for TEM irradiated NVs located 50 nm below the surface of the diamond. Note that multiple NVs were found with $T_2$ values greater than or equal to 1 ms and all $T_2$ values are longer than 100 μs.

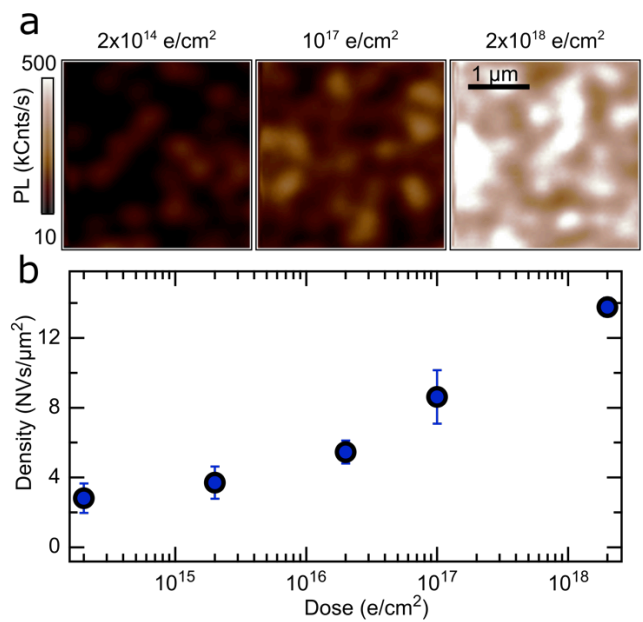

**Figure 3.** **(a)** Confocal photoluminescence images of diamond areas irradiated with 200-keV incident energy electrons and doses ranging from $2 \times 10^{14}$ e/cm$^2$ to $2 \times 10^{18}$ e/cm$^2$. The color scale bar is the same for every image to more easily compare the PL of the surface. **(b)** Plot of NV density versus irradiation dose. The error bars show the goodness of fit for the maximum likelihood estimate model to the data.

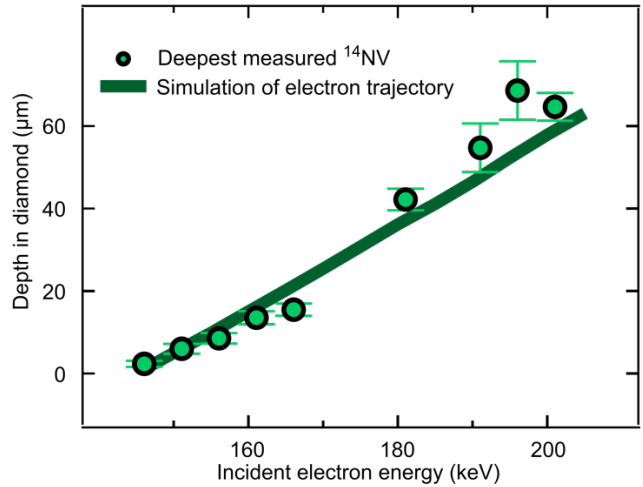

**Figure 4.** Depth of deepest NV formed via TEM irradiation in areas irradiated by incident electrons ranging in energy from 145 keV to 200 keV. Data are green circles. Simulation results (green line) model the travel depth of electrons before reaching a threshold energy of 145 keV. The slope of the simulation (change in depth for incident electron energy) is 1.05 ± 0.01 $\mu$m/keV. The slope of the weighted fit to the data (line not shown) is 1.05 ± 0.04 $\mu$m/keV: increasing the electron energy by 1 keV results in the formation of NVs 1 $\mu$m deeper in the diamond. Hence NV creation is an excellent marker for vacancy formation.

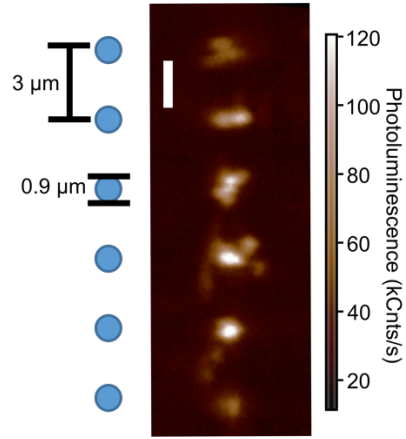

**Figure 5. Lateral position control of NV formation.** Photoluminescence image showing NV centers formed in the TEM irradiation spots with ~ 1-μm lateral control. The scale bar in the image is 3 μm. The pattern made with the TEM is shown on the left.